# Identity Ordering and Metastable Clusters in Fluids with Random Interactions

Itay Azizi and Yitzhak Rabin

Department of Physics, and Institute of Nanotechnology and Advanced Materials, Bar-Ilan University, Ramat Gan 5290002, Israel

**Abstract**

We use Langevin dynamics simulations to study dense two-dimensional systems of particles where all binary interactions are different (AID) in the sense that each interaction parameter is characterized by a randomly chosen number. We compare two systems that differ by the probability distributions from which the interaction parameters are drawn: uniform (U) and exponential (E). Both systems undergo neighborhood identity ordering (NIO) and form metastable clusters in the fluid phase near the liquid-solid transition but the effects are much stronger in E than in U systems. Possible implications of our results for the control of the structure of multicomponent alloys are discussed.

## I. Introduction

Systems in which all particles are different (APD) and those in which all interactions $\epsilon_{ij}$ between particles are different (AID) sparked the interest of theoretical physicists as they can serve as a generic model for extremely heterogeneous systems which abound in material science and biology [1-6]. Molecular dynamics simulations of both high density [7-9] and low density [10] Lennard-Jones fluids (and also Monte Carlo simulations [11]) have been performed to characterize the behavior of such systems.

It was shown that simple principles guide their local and global organization and their dynamics: particles pick neighbors to minimize the potential energy of the system (neighborhood identity ordering, NIO), resulting in microphase separation in the fluid phase that becomes increasingly pronounced as one approaches the liquid-solid transition.

In the following we assume that particles interact via Lennard-Jones forces, with the same size parameter and with pair interaction parameters $\epsilon_{ij}$ that are drawn at random from some given distribution. The polydispersity degree is defined as

$$\Delta = \frac{SD}{<\epsilon_{ij}>} \quad (1)$$

where SD stands for the standard deviation of $\epsilon_{ij}$. This parameter shows how broad the distribution is compared to its mean. For one-component (1C) systems the polydispersity degree is zero while for the uniform distribution in the range $1 \leq \epsilon_{ij} \leq 3$, $P_0^U(\epsilon_{ij}) = 1/2$ (independent of $\epsilon_{ij}$), $<\epsilon_{ij}>_0 = 2$ and $\Delta \approx 0.29$. Since the distributions considered in our previous work [7], 1C and U, were quite narrow, in this work we decided to study the exponential distribution which has a very broad range $[1, \infty]$ of interaction parameters $\epsilon_{ij}$ between a pair of particles i and j (Fig. 1). The normalized exponential distribution with mean $<\epsilon_{ij}>_0 = 2$ is given by

$$P_0^E(\epsilon_{ij}) = e^{-\epsilon_{ij}+1} \quad (2)$$

Compared to the uniform (U) distribution in which half of the bonds $\epsilon_{ij}$ (we will use the term "bond" and "interaction parameter" interchangeably in this work) are weaker than average, $<\epsilon_{ij}>_0 = 2$, the exponential (E) distribution is enriched with weak bonds (~63% of all bonds) and has a larger polydispersity degree, $\Delta = 0.5$. Nevertheless, even though only a smaller fraction of the N(N-1)/2 bonds in E are strong (N is the number of particles in the system), the fact that in each configuration of a dense system in 2d there are about 3N bonds (assuming that most particles have 6 nearest neighbors), means that at sufficiently low temperatures energy minimization will drive the system towards configurations in which the majority of the bonds present are strong and therefore the distribution of bonds in such configuration will strongly deviate from the random one given in Eq. (2). This is achieved by neighborhood identity ordering (NIO) - particles that strongly attract each other tend to remain close and form transient clusters. The study of such effects is the aim of the present work.

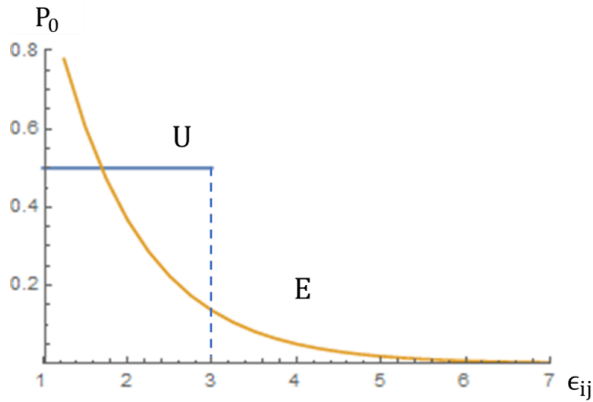

Fig. 1 Probability distributions of the interaction parameters: uniform (U), $P_0^U = 1/2$ and exponential E, $P_0^E = e^{1-\epsilon_{ij}}$. The average of both distributions is $<\epsilon_{ij}> = 2$.

In section II we present the computational model and discuss the simulation algorithm and the various parameters. In section III we present the results of our computer simulations and compare the E system to U and the one-component (1C) systems. In section IV we discuss the main results of this work and the new insights obtained about the physics of fluids with exponentially and uniformly distributed interactions.

## II. Model and Simulation Details

We perform Langevin dynamics simulations using LAMMPS (in NVT ensemble) in two dimensions with N=2500 particles in a simulation box of size $L_x = L_y = 60\sigma$ (where $\sigma$ is the particle diameter), at triple point density $0.69\sigma^{-2}$ (=$N/L_x L_y$). Periodic boundary conditions are applied along both x and y directions. All particles have same size $\sigma$ and mass m which will be set to unity. The particles interact via Lennard-Jones (LJ) potential

$$V_{ij}(r) = 4\epsilon_{ij}[(\sigma/r)^{12} - (\sigma/r)^6] \qquad (3)$$

where $\epsilon_{ij}$ and r are the pair interaction parameter and separation respectively, between particles i and j. The potential is truncated and shifted to zero at $r = 2.5\sigma$ (the small discontinuity of the force at the cutoff distance does not affect our results since its magnitude is very small compared to the thermal force $k_B T/\sigma$).

The equation of motion of particle i is given by the Langevin equation

$$m\frac{d^2 r_i}{dt^2} + \zeta \frac{dr_i}{dt} = -\frac{\partial V}{\partial r_i} + \boldsymbol{f_i} \quad (4)$$

with $r_i$ the position of particle i, $\zeta$ the friction coefficient (assumed to be the same for all particles), V is the sum of all pair potentials $V_{ij}$ (i ≠ j) for a given spatial configuration of the system, and $\boldsymbol{f_i}$ is a random force with zero mean and second moment proportional to the product of temperature T and $\zeta$. All physical quantities are expressed in LJ reduced units, with LJ time $\tau_{LJ} = 1$ (the simulation time step used in the integration of the equations of motion is $0.005\tau_{LJ}$). The friction coefficient can be written as $\zeta = 1/\tau_d$, with $\tau_d$ the characteristic viscous damping time which determines the transition from inertial to overdamped motion (due to collisions with molecules of the implicit "solvent"). The viscous damping time was taken to be the same as in our previous studies of APD and AID systems [7,8,10], $\tau_d = 50\tau_{LJ}$.

In order to characterize the interactions between the different particles, we have to specify the distribution $P_0(\epsilon_{ij})$ of the pair interaction parameters that enter the LJ potential in Eq. 3. The subscript 0 in $P_0(\epsilon_{ij})$ stands for the intrinsic distribution from which all the possible N(N-1)/2 bonds in the system are drawn. In general, this distribution is different from the distribution $P(\epsilon_{ij})$ of the values of $\epsilon_{ij}$ present in any particular configuration (snapshot) of the system. $P(\epsilon_{ij})$ depends on the identity of the nearest neighbors of each of the particles in this configuration and approaches $P_0(\epsilon_{ij})$ only in the high temperature limit where the fluid is highly disordered and all configurations are equally probable. In this work we compare two intrinsic distributions (Fig. 1): (1) Uniform (U) distribution, in which each pair of particles i and j is assigned an interaction parameter $\epsilon_{ij}$ taken from a uniform distribution in the range from 1 to 3. (2) Exponential (E) distribution given in Eq. 2, where the $\epsilon_{ij}$ are drawn from an exponential distribution in the range from 1 to infinity. To facilitate the comparison between the two distributions, the width of the exponential distribution is chosen such that its mean $<\epsilon_{ij}>_0 = 2$ coincides with that of the uniform distribution.

According to the phase diagram of one-component (1C) LJ systems in 2 dimensions (Fig. 1 in Ref. 12), the freezing transition takes place at $T^* = 0.4\ \epsilon_{ij}$. Taking the interaction parameter to be equal to the average value of both the U and the E distributions yields $T^* = 0.4 \times 2 = 0.8$. The exact nature of this transition is still under dispute [13]. According to theoretical arguments [14] and some simulations [15] freezing is a two-step process: a continuous transition from normal to hexatic liquid followed by another continuous transition from hexatic liquid to solid. Other researches claim that it is a first order transition, possibly broadened by finite size effects in computer simulations [16].

We carried out simulations of 1C and U systems using the following method: we begin from a square lattice configuration and equilibrate a fluid at a sufficiently high temperature (at which it is nearly completely disordered), and then cool the system to the target temperature until steady state is reached in the sense that ensemble-averaged properties such as the mean potential energy per particle, become time-independent and no further aging is observed. While for 1C and U systems it is sufficient to start at T=3 fluid, the presence of very high $\epsilon_{ij}$ bonds ($\epsilon_{ij} \gg 2$) in the E system requires us to go to higher temperatures in order to achieve equilibration in respect to the formation and breakup of all observed bonds. To this end we begin all the simulations of the E system at T=10.

In all three systems (1C, U and E), equilibration at the final temperature is quite fast sufficiently far above the transition but becomes very slow as the transition is approached. For example, just above the transition the relaxation time is about 2,500 $\tau_{LJ}$ for 1C, 10,000 $\tau_{LJ}$ for U and about 50,000 $\tau_{LJ}$ for the E system. Even slower relaxation is observed at and below the transition and while the one-step cooling procedure is sufficient for the 1C and the U systems, in order to approach steady state at these temperatures for the E system, we used a two-step temperature quench:

1. Cool from equilibrated T=10 fluid to T=1.3 and equilibrate the system at this temperature.

2. Cool from T=1.3 fluid to the target temperature and let the system relax to steady state at this temperature.

While equilibration of the E system at the final temperature was always achieved above the liquid-solid transition, cooling below the transition resulted in fast relaxation followed by much slower approach to equilibrium which continued through the longest times reached in our simulations (not shown).

## III. RESULTS

In order to determine the freezing transition temperatures of 1C, U and E systems we proceeded to measure the mean potential energy per particle in steady-state at each temperature (Fig. 2). Note that while the potential energy of the U system is lower than that of the 1C system, the E system exhibits much more dramatic lowering of the potential energy and its freezing transition is shifted to much higher temperature. This behavior is the result of enhanced sampling of strong bonds, which is especially pronounced in the E system because of the exponential tail of high $\epsilon_{ij}$ bonds. From the potential energy we proceed to calculate the specific heat as function of temperature,

$$c_v = \frac{dU}{dT} = \frac{d(e_p + e_k)}{dT} = \frac{de_p}{dT} + 1 , \tag{5}$$

where U is the total energy per particle, $e_p$ is the potential energy per particle and $e_k = T$ is the kinetic energy per particle.

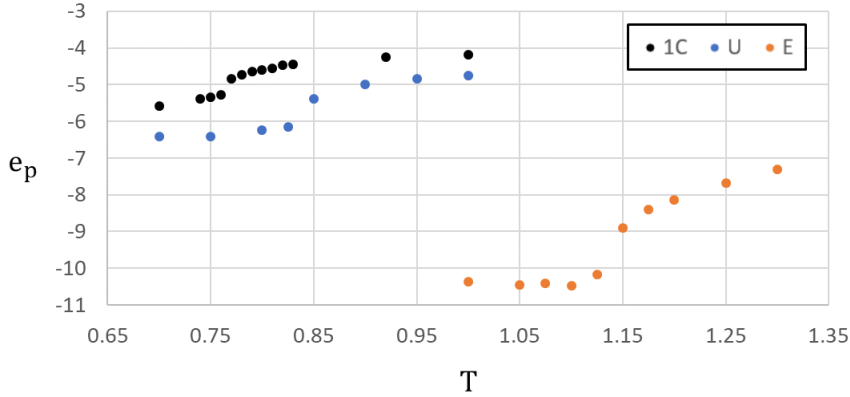

Fig. 2  Mean potential energy per particle as function of temperature.

Next we measured the mean hexatic order parameter $< X_6 >$, where $X_6$ is the hexatic order parameter of a single configuration (snapshot) defined as:

$$X_6 = \left| \frac{1}{N} \sum_{j=1}^{N} \psi_6^j \right| \tag{6a}$$

$$\psi_6^j = \frac{1}{n_j} \sum_{k=1}^{n_j} e^{i6\theta_{jk}} \tag{6b}$$

N is the number of particles, $n_j$ is the number of nearest-neighbors of particle j, the sum is over its neighbors j and $\theta_{jk}$ is the angle between an arbitrary fixed axis and the line connecting particles j and k.

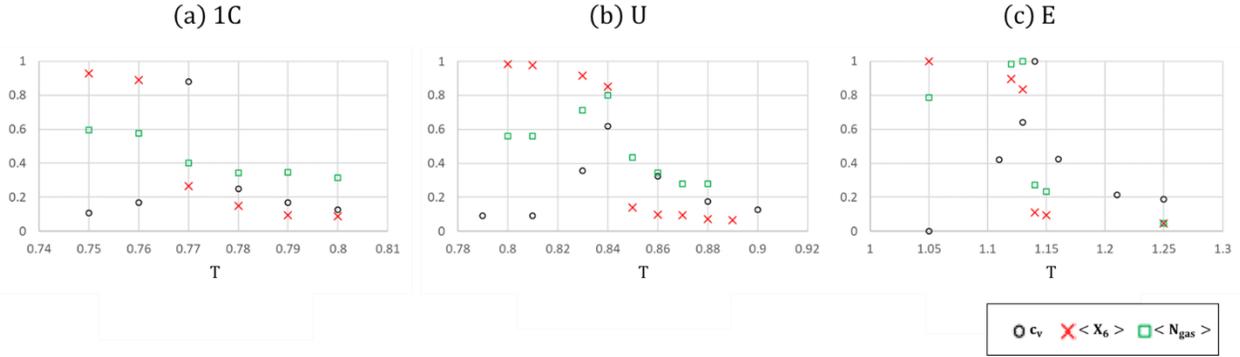

Fig. 3 Specific heat, mean hexatic order parameter and mean number of gas particles as function of temperature (in arbitrary units in the range 0-1).

Finally, since the transition is between a homogeneous liquid and a phase in which solid and gas phases coexist, we measured the mean number of isolated particles inside the voids in the surrounding dense phase. In Fig. 3 we plot the 3 quantities, the specific heat, the average hexatic order parameter and the average number of isolated particles in the gas phase (all of them in arbitrary units, in the range 0 to 1) as a function of temperature, for the 3 systems. The transition temperatures were identified from the peak in the specific heat and from the crossover between the limiting values of the mean hexatic order parameter and the number of isolated particles: $T_{1C}^* \cong 0.77$, $T_U^* \cong 0.84$ and $T_E^* \cong 1.14$. Note that the value 0.77 is somewhat lower than the reported equilibrium transition temperature (0.8) of the two-dimensional 1C LJ system [12], presumably due to our cooling procedure and finite system size effects. The relatively high transition temperature of the E system is due to the effect of the high $\epsilon_{ij}$ tail of this distribution that increases the temperature at which energy and entropy balance each other, compared to the U system. In all the systems, we observe a single transition characterized by the appearance of hexatic order and without visible intermediate hexatic liquid phase above the transition.

In order to compare the three systems at the same distance from their respective transition temperatures we introduce the reduced temperature

$$\delta = \frac{T-T^*}{T^*} \tag{7}$$

In Fig. 4 we show snapshots of the different systems at two different temperatures above the transition. In order to display the degree of neighborhood identity ordering (NIO) in the U and the E systems we define the effective interaction parameter of particle i as an average over the interaction parameters with its $n_i$ nearest neighbors

$$\epsilon_i^{eff} = \frac{1}{n_i}\sum_{j=1}^{n_i}\epsilon_{ij} \tag{8}$$

and color the particles according to their $\epsilon_i^{eff}$ values. As shown in Fig. 4, at $\delta = 0.1$ above the freezing transition, there is visible NIO in both U and E (but of course, not in 1C) but the aggregation of particles with high $\epsilon_i^{eff}$ values is much larger in E than in U, presumably due to the high $\epsilon_{ij}$ tail of the exponential distribution. Small, rapidly moving vacancies are present in the three systems. Closer to the transition, at $\delta = 0.02$, NIO is strongly enhanced in the E system and less so in the U system. The voids increase in size by coalescence as the transition is approached, especially in the 1C and the U systems.

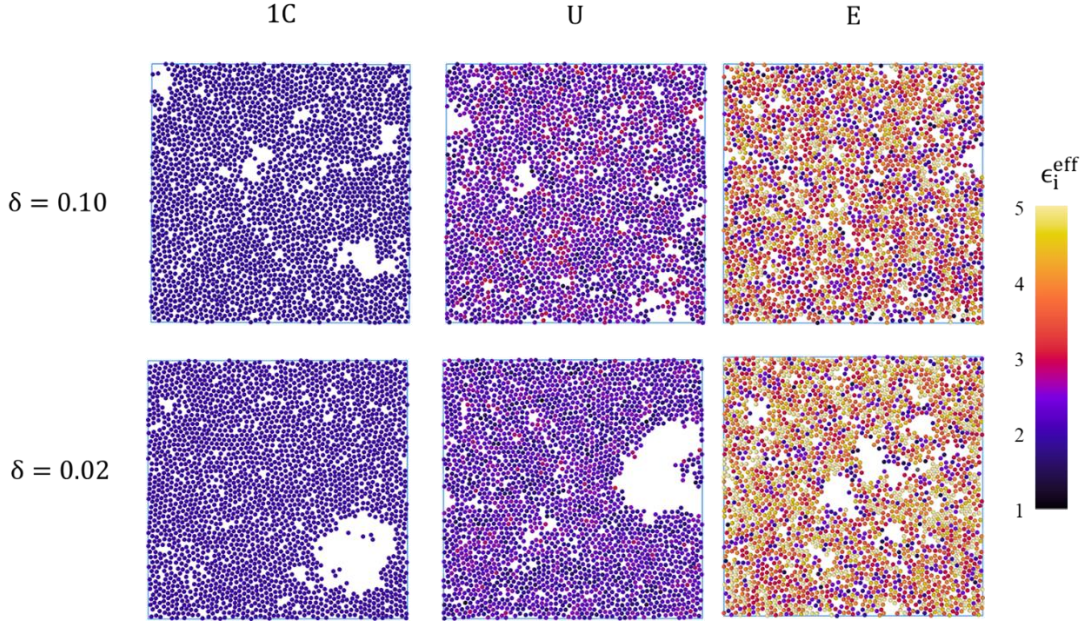

Fig. 4 Snapshots of the systems in steady-state, at $\delta = 0.10$ and $\delta = 0.02$ (particles with $\epsilon_{eff} \geq 5$ are color-coded as 5).

In Fig. 5 we present the mean interaction parameter $<\epsilon_{ij}>$ as a function of the reduced temperature above the transition. In both the U and the E systems this parameter is a monotonously decreasing function of temperature but the values of $<\epsilon_{ij}>$ are much higher in the E than in the U system. Note that while in the high T limit (outside the range of temperatures shown in Fig. 5), the mean interaction parameter of both systems must approach the average of the intrinsic U and E distributions $<\epsilon_{ij}>_0 = 2$, it takes a higher temperature in the E system to reach a truly disordered state in which NIO is completely suppressed.

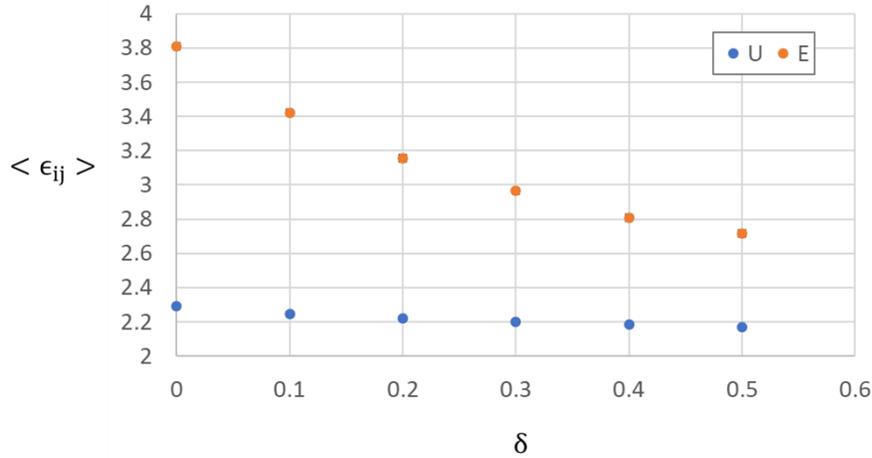

Fig. 5 Mean interaction parameter as a function of reduced temperature (error bars are too small to be seen).

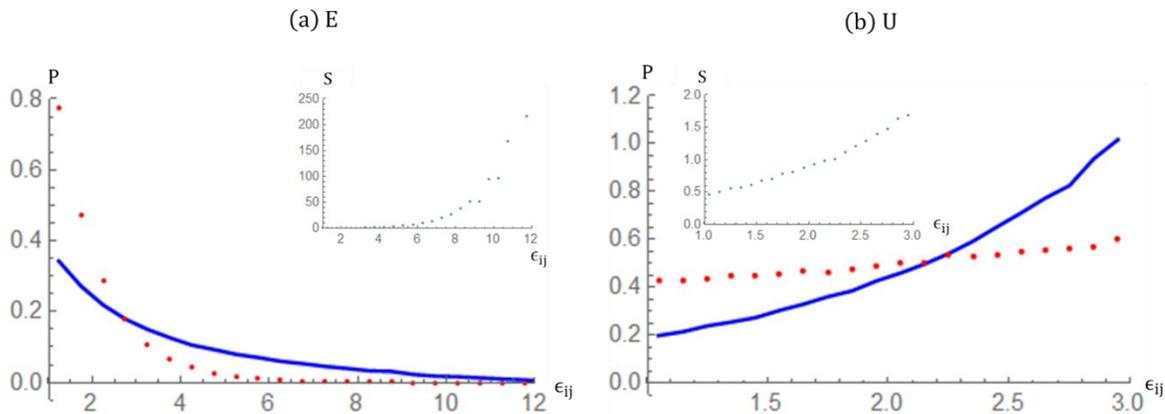

Fig. 6 High temperature (red dotted line) and $\delta = 0.01$ (blue solid line) $\epsilon_{ij}$ distributions for (a) E and (b) U systems. The ratio S of the low temperature to the high temperature distribution is shown in the insets. The averages of the distributions shown are
(a) $<\epsilon_{ij}>_{T=1.15} = 3.76$, $<\epsilon_{ij}>_{T=10} = 2.03$, (b) $<\epsilon_{ij}>_{T=0.85} = 2.27$, $<\epsilon_{ij}>_{T=3} = 2.06$.

In order to get an intuitive feeling about the magnitude of NIO in the E and the U systems as one approaches the freezing transition, we compare the interaction parameter distribution $P(\epsilon_{ij})$ of the high temperature fluid to the corresponding distribution slightly above the freezing transition temperature (Fig. 6). The ratio S of the distributions at $\delta = 0.01$ and at high temperature (T=10 for E and T=3 for U, respectively) is shown in the insets to the figures. The dramatic enhancement of the high $\epsilon_{ij}$ tail due to NIO is clearly observed in the E system. Enhancement of the large $\epsilon_{ij}$ part of the distribution is also observed in the U system but the effect is much smaller than in E (note the different scales on the y axes of the insets of Figs 6a and 6b). As can be seen from the inset of Fig. 6a, the sampling of $\epsilon_{ij} \geq 10$ in the E system is 150 times larger at $T = 1.15$ than at T=10.

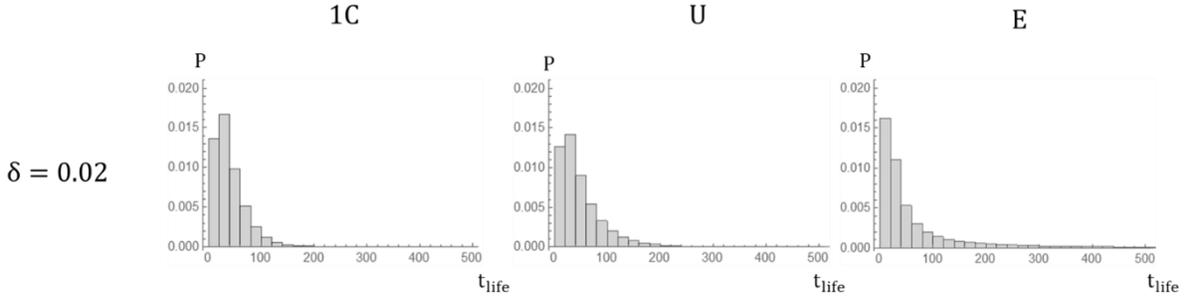

Fig. 7 Histograms of bond lifetime distributions at $\delta = 0.02$ (lifetimes are in $\tau_{LJ}$ units). The measurement interval is $\Delta t = 5000\ \tau_{LJ}$.

We now proceed to examine the dynamics of uniform and exponential systems near the freezing transition by measuring the lifetime of individual bonds and that of large connected clusters as a function of temperature, for the E, U and 1C systems. Here, the lifetime of a bond is defined as the time interval during which the inter-particle distance remains below $1.7\sigma$ (the location of the minimum of the pair correlation function [17]). To this end, we follow all the pair interactions in the system within a time interval of $5000\ \tau_{LJ}$ and measure the lifetimes of the corresponding bonds (the error of such a measurement is $5\ \tau_{LJ}$). As shown in Fig.7, above the transition ($\delta = 0.02$) the bond lifetime distribution function is a monotonically decreasing function of lifetime, but with different averages for 1C, U and E systems: $\langle t_{life} \rangle$ =37 $\tau_{LJ}$ for 1C, 47 $\tau_{LJ}$ for U and 197 $\tau_{LJ}$ for E (the difference is due to the presence of a long tail in the E distribution of lifetimes and of bond strengths).

We expect the lifetime of a bond ($t_{life}$) to be strongly correlated with its strength ($\epsilon_{ij}$). This is clearly observed in Fig. 8 where we plot the $\epsilon_{ij}$ distributions corresponding to 3 ranges of lifetimes above the transition ($\delta = 0.1$). For short-lived bonds, the corresponding distributions are close to exponential (for E) and uniform (for U) and their means $\langle \epsilon_{ij} \rangle$ are close to 2. For longer-lived bonds the distributions deviate dramatically from their intrinsic forms (exponential and uniform, respectively) and $< \epsilon_{ij} >$ increases with $t_{life}$.

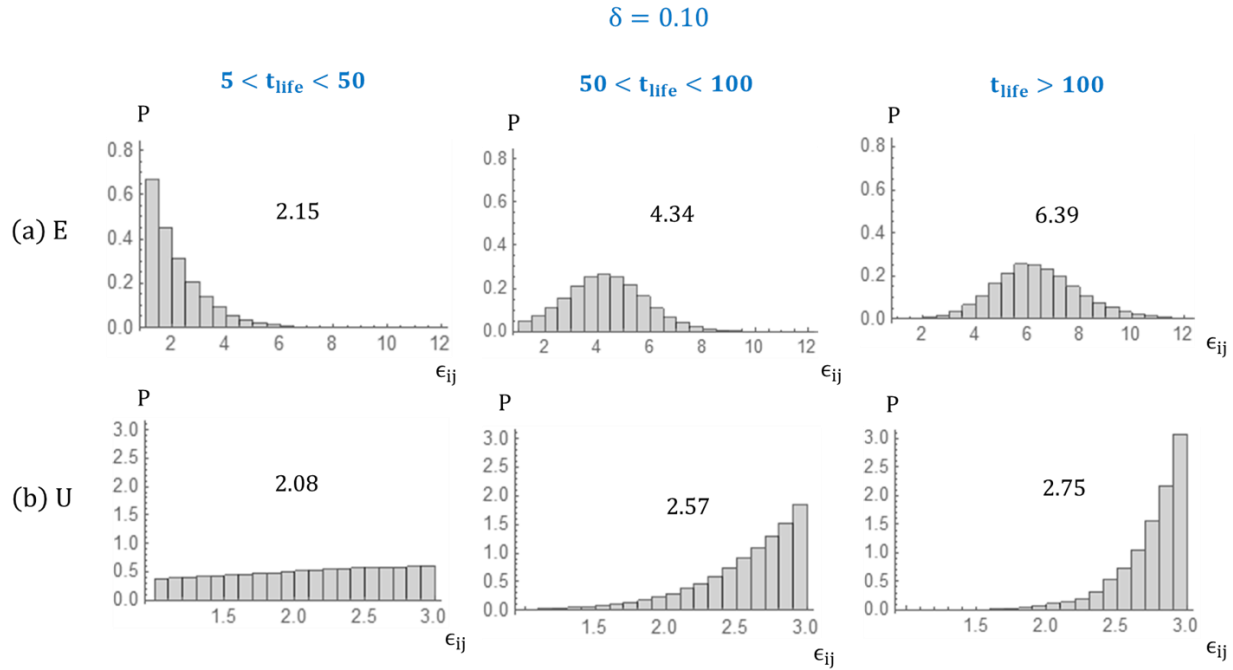

Fig. 8 Histograms of $\epsilon_{ij}$ distributions in different ranges of lifetimes (indicated above in $\tau_{LJ}$ units) at $\delta = 0.10$. The average of the distribution is shown on each histogram.

After addressing the lifetime distributions, the next question is what is the number of long-lived bonds in a given configuration of the system and how does it vary with temperature. Note that all bonds become short-lived in the high temperature limit: a typical bond in the U system at T=3 or in the E system at T=10, has an average lifetime of about 10 $\tau_{LJ}$ (not shown). In the following we introduce a somewhat arbitrary cutoff time such that bonds with lifetimes equal or larger than $t_{cutoff}$ ($\gg 10\ \tau_{LJ}$) are considered as long-lived bonds (LLB). As shown in Fig. 9 for $t_{cutoff} = 100\ \tau_{LJ}$, the mean number of LLB is a monotonously decreasing function of temperature and is the highest for E, lower for U and lowest for 1C. Similar behavior is observed for other cutoff times, 50 $\tau_{LJ}$ and 75 $\tau_{LJ}$ (not shown).

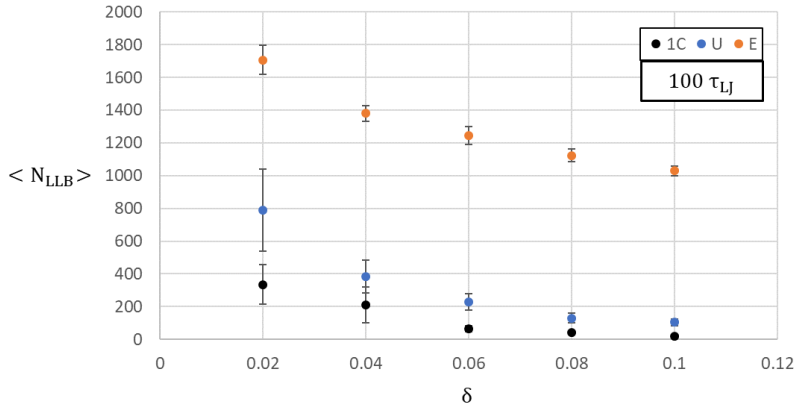

Fig. 9 Mean number of long-lived (longer than $100\ \tau_{LJ}$) bonds as a function of reduced temperature.

Long-lived bonds can form large metastable structures (clusters). A cluster is defined as a set of LLB with a lifetime that exceeds a certain cutoff time, which form a connected graph such that all the components of the cluster (its bonds) remain intact during $t \geq t_{cutoff}$. We measured the number of particles $M_{LLB}$ in the largest cluster in each configuration of the system and calculated its average (over independent snapshots) as a function of temperature (Fig. 10a). We find that for $t_{cutoff} = 100\ \tau_{LJ}$, the mean size of the largest cluster decreases with increasing temperature and is always higher in E than in U. Notice that 1C clusters are not observed for this cutoff time because the number of LLB in the system (Fig. 9) is too low for such clusters to form.

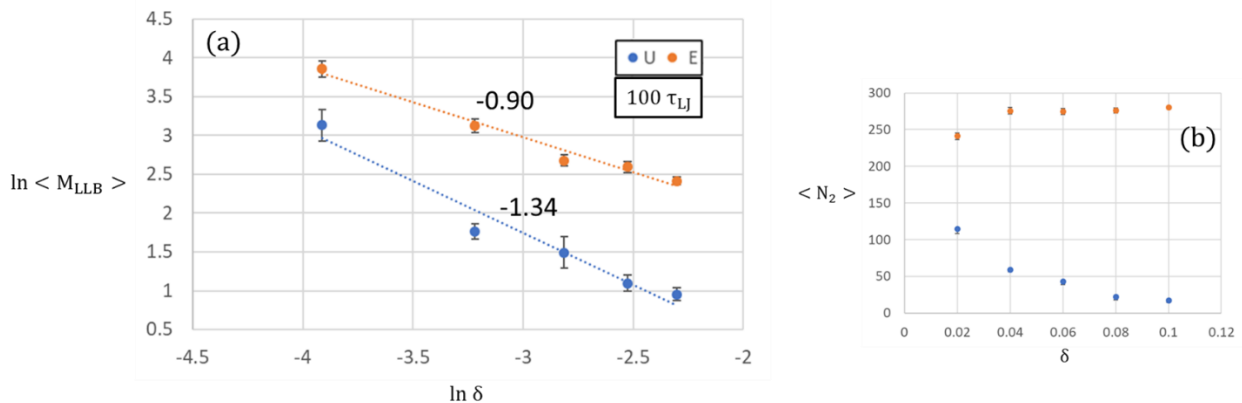

Fig. 10 (a) ln-ln plot of the average size of the largest LLB cluster as function of reduced temperature. (b) Mean number of LLB diatomics as function of reduced temperature (both are composed of bonds which live longer than $100\ \tau_{LJ}$).

The temperature dependence of the mean size of the largest cluster follows a power law (Fig. 10a)

$$< M_{LLB} > \sim \delta^{-\alpha}. \tag{9}$$

For the E system, the exponent α is nearly independent of the cutoff time: for $t_{cutoff} = 100\,\tau_{LJ}$, $75\,\tau_{LJ}$, $50\,\tau_{LJ}$ we get $\alpha_E = 0.90, 0.88$ and $0.88$, respectively. For the U system α exhibits non-monotonic dependence on the cutoff time: for $t_{cutoff} = 100\,\tau_{LJ}$, $75\,\tau_{LJ}$, $50\,\tau_{LJ}$ we get $\alpha_U = 1.34, 1.51$ and $1.47$, respectively. The average bond strength $<\epsilon_{ij}>_C$ of bonds that belong to a cluster (Fig. 11) is larger than that of all of the bonds $<\epsilon_{ij}>$ at the same temperature, in accord with the expectation that metastable structures are formed by stronger bonds.

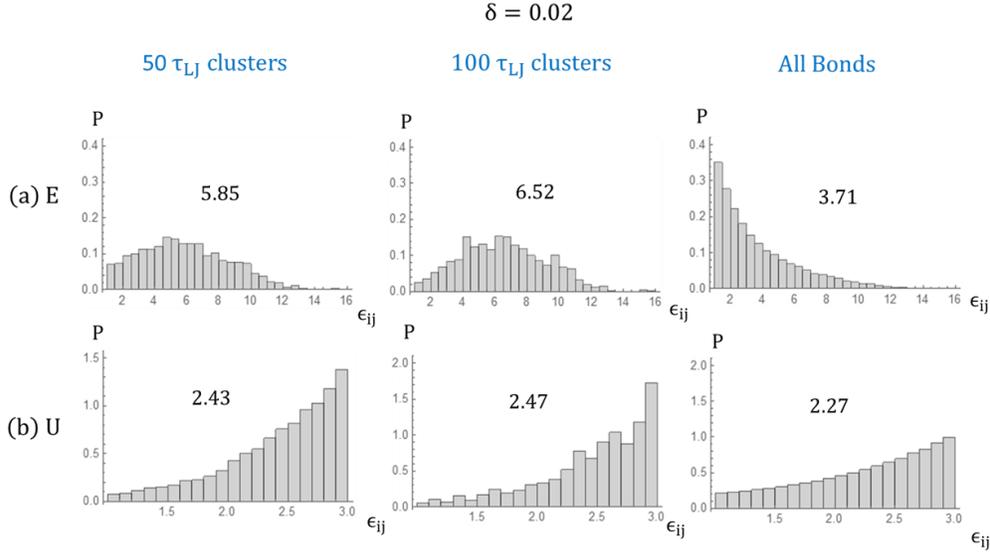

Fig. 11 Histograms of $\epsilon_{ij}$ distributions of the bonds of the largest cluster (according to time criterion written above) and of all bonds in the system, at $\delta = 0.02$ (the average is indicated on each histogram).

It is also interesting to consider the smallest clusters -i.e., isolated pairs of particles connected by a single long-lived bond. In Fig. 10b we plot the mean number of long-lived pairs $<N_2>$ as a function of temperature. The number such pairs is always much larger in E than in U, due to the contribution of high $\epsilon_{ij}$ bonds that are present in the exponential but not in the uniform distribution. The observation that the number of long-lived pairs in the U system decreases with increasing temperature concurs with our intuitive expectations based on the Boltzmann distribution.

Surprisingly, the number of long-lived pairs in the E system is nearly temperature-independent above the transition (it decreases at yet higher temperatures not shown in Fig. 10b) but goes down as the freezing transition is approached. Upon some reflection we conclude that this is the consequence of the incorporation of these pairs into large metastable clusters that form near the transition temperature (only isolated pairs contribute to $<N_2>$).

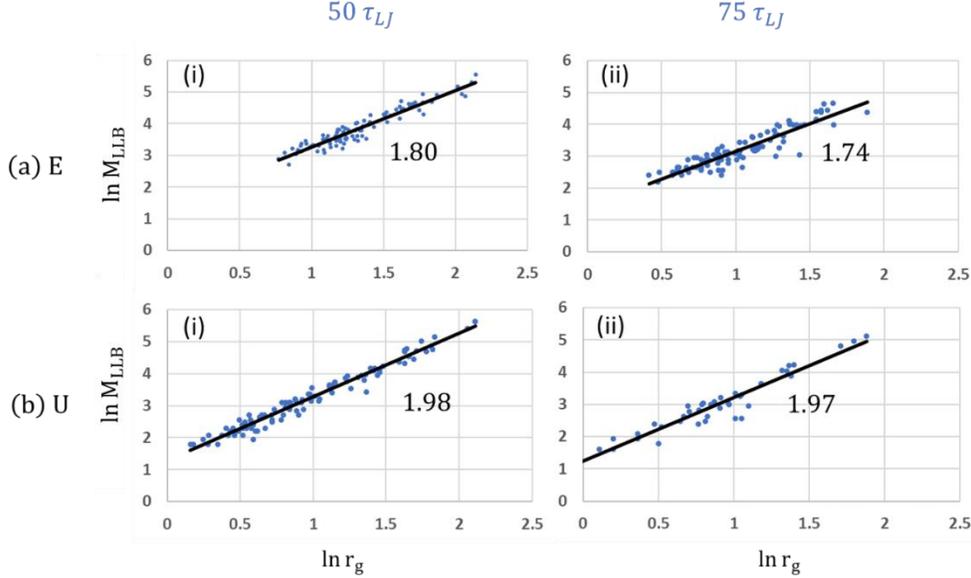

Fig. 12 ln-ln plot of number of particles per cluster as function of radius of gyration of the largest long-lived (according to time criterion written above) clusters in the temperature range $\delta = 0.02 - 0.10$ (for U bonds with lifetime longer than $75\tau_{LJ}$ only clusters in the temperature range $\delta = 0.02 - 0.04$ are shown because above this temperature range the clusters are too small). The best fit to the slope is shown on each plot.

In order to gain further insight about the shape of the metastable clusters we calculated the fractal dimension of the clusters $d_f$ using the standard definition of the Hausdorff dimension,

$$M_{LLB} \sim r_g^{d_f} \qquad (10)$$

where $r_g$ is the radius of gyration of the cluster. For $t_{cutoff} = 100\ \tau_{LJ}$ U clusters are too small to be characterized by a fractal dimension and, therefore, we only calculated their fractal dimensions for lower cut-off times. We find that typical U clusters are compact with $d_f \simeq 2$ and that E clusters are somewhat less compact, with $d_f$ in the range $1.75 - 1.80$ (Fig. 12). These estimates are consistent with visual examination of snapshots of the clusters (Figs. 13-14 and Supplementary Material (1) movie of E clusters) and can be rationalized as follows. Since the U system does not contain high $\epsilon_{ij}$ bonds, U clusters maintain stability (form LLB) by condensation that increases the number of nearest-neighbors and results in compact structure. Conversely, E clusters can be stabilized by a small number of strong (high $\epsilon_{ij}$) bonds resulting in ramified structures with lower fractal dimension.

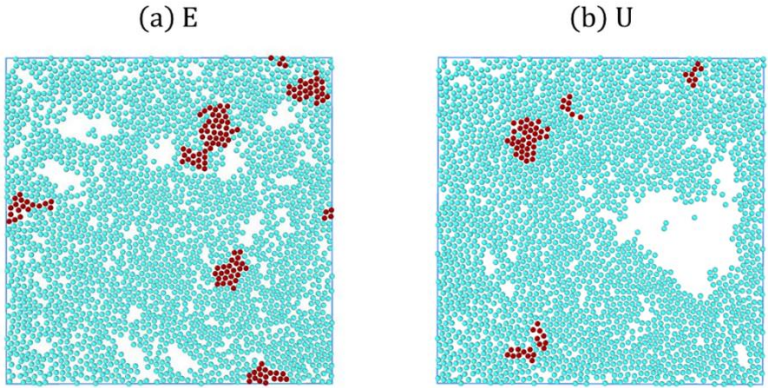

Fig. 13 System snapshots at $\delta = 0.02$. 5 largest clusters ($t_{cutoff} = 100\ \tau_{LJ}$) are shown in brown.

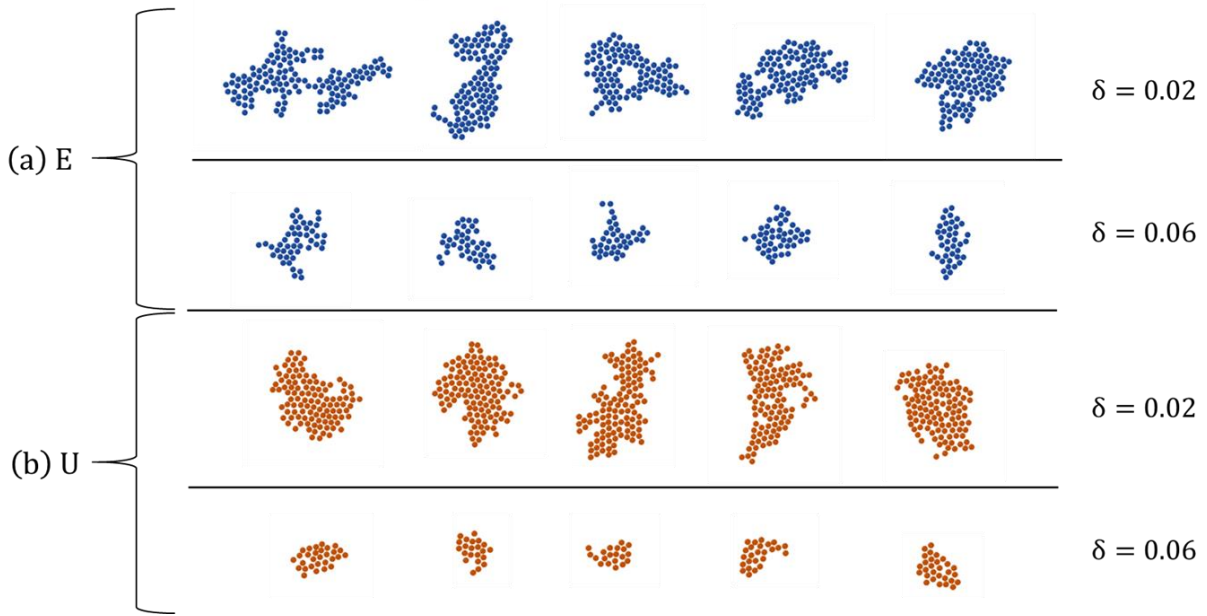

Fig. 14 Snapshots of long-lived clusters ($t_{cutoff} = 50\ \tau_{LJ}$) at $\delta = 0.02$ and $\delta = 0.06$ (each cluster is separated by at least $1000\ \tau_{LJ}$ from its neighbour).

## IV. Discussion

Both static and dynamic properties of fluids with random interactions depend on the statistical distributions from which the pair interaction parameters (bonds) are drawn at random. In our previous studies we considered the uniform (U) distribution in which all possible values of the pair interaction parameters in a given range [$\epsilon_{ij}^{min}$, $\epsilon_{ij}^{max}$] are equally probable. In the present work we compared the U distribution to the exponential (E) distribution in which no upper limit was set for $\epsilon_{ij}$ (we took $\epsilon_{ij}^{min} = 1$ for both distributions). Our choice of the exponential distribution was motivated by a fundamental difference between U and E: while in the U distribution $\epsilon_{ij}^{min}$ and $\epsilon_{ij}^{max}$ are equally probable, in the E distribution the probability of generating high $\epsilon_{ij}$ values is exponentially small compared to that of $\epsilon_{ij}^{min}$. The combination of a very broad range of accessible $\epsilon_{ij}$ values with a relatively small value of the mean of the distribution (we took $<\epsilon_{ij}> = 2$ for both distributions thus fixing $\epsilon_{ij}^{max}(U) = 3$) is a characteristic of the exponential but not of the uniform distribution. Since preliminary results indicate that in this respect power law distributions behave similarly to the U distribution, we decided not to include them in this study.

The most important feature of fluids with random interactions is the appearance of a broad range of temperatures in which there is a new type of order, neighborhood identity ordering (NIO). While in the high temperature limit the probability to find a bond of strength $\epsilon_{ij}$ in any snapshot (instantaneous state) of the system is given by the uniform or the exponential distribution for the U or the E system, respectively, at lower temperatures this probability is affected by the increased weight of low energy configurations. As temperature is lowered, the system reduces its energy by rearranging the particles into local neighborhoods in which particles that strongly attract each other are concentrated, forming microphase separated patterns. Since the number of possible bonds between N particles is N(N-1)/2 while each instantaneous configuration of the system has only less than 3N bonds, this rearrangement corresponds to selection of higher values of interaction parameters and results in a new, temperature-dependent distribution in which high $\epsilon_{ij}$ values are enhanced and low $\epsilon_{ij}$ values are suppressed compared to the high temperature distribution. At temperatures above the freezing transition the states of the system are highly degenerate in the sense that a very large number of configurations (each configuration is characterized by a particular set $\{\epsilon_{ij}\}$) yields the same total energy of the system. The fact that in the fluid phase the system is able to move between states that correspond to different sets of bonds $\{\epsilon_{ij}\}$ by local rearrangement of the particles means that NIO is compatible with fluidity and that we have a liquid in which there exist strong local correlations between particle identities. Of course, at yet lower temperatures the system freezes and no further rearrangement (and therefore no further enhancement of NIO) is possible.

In the present study we have shown that the above mentioned effects are dramatically enhanced in the E compared to the U system. Thus, while NIO as measured by $<\epsilon_{ij}>$ increases only by 10% on cooling the U system from T=∞ to the freezing transition, it nearly doubles in the E system (Fig. 5). The effect of NIO on the tails of the distributions is even more dramatic; as one approaches the transition, the enhancement of the tail of the E distribution (i.e., of $\epsilon_{ij} \geq 10$) is several orders of magnitude larger than that of $\epsilon_{ij}^{max} = 3$ of the U distribution.

The presence of the high $\epsilon_{ij}$ tail and the resulting enhancement of NIO in the E compared to the U system, is reflected in the different dynamics of the two systems. Thus, typical equilibration times (not shown) and lifetimes of bonds (Fig. 7) are always larger in E than in U system, due to the presence of high energy bonds in the former (Fig. 8). Long-lived bonds proliferate with the approach to the freezing transition and their number is much greater in the E than in the U system (Fig. 9). The relative abundance of long-lived bonds in the E compared to the U system is even more pronounced at higher temperatures due to the contribution of high $\epsilon_{ij}$ bonds in the former which can form metastable "diatomics" even quite far from the transition (Fig. 10b). At lower temperatures large metastable clusters that grow in size as the transition is approached appear in both U and E (but not in the 1C) system (Fig. 10a). These clusters are larger in E than in U and while the latter are compact, the E clusters are ramified with a fractal dimension of about 1.8 (Figs. 12-14). Both U and E clusters are stabilized by the fact that each particle in the bulk of a compact solid-like cluster is confined to a "cage" formed by its neighbors. The ramified structure of the E clusters can be traced back to the presence of very high $\epsilon_{ij}$ bonds (Fig. 11) that can remain connected to the rest of the cluster for the duration of its lifetime, even without such caging effects.

We would like to mention that the appearance of large solid-like clusters in fluids with random interactions near their freezing temperature, is a characteristic signature of criticality. This is quite unexpected since in one-component systems critical behavior is associated with the gas-liquid critical point but not with liquid-solid transitions [15]. It has been recently pointed out that dynamic heterogeneities near the glass transition in size-polydisperse Lennard-Jones fluids are strikingly similar to critical fluctuations near a critical point [16]. While in our size-monodisperse AID systems such a glass transition is preempted by crystallization, there are indications of a glass transition with respect to NIO due to slowing down of the kinetics of particle rearrangement [8]. Whether the freezing of AID fluids in two dimensions is a true critical phenomenon cannot be answered by the present study and requires further study.

We would like to comment on a possible application of the present study to multicomponent alloys [20, 21]. Since AID/APD fluids become progressively ordered as the fluid is cooled above the freezing temperature, this suggests that one can manipulate the structure of multicomponent alloys by first equilibrating the system at a high temperature liquid state, and then performing a rapid quench to room temperature. Depending on the initial temperature one can then obtain a homogeneous solid "solution" by first equilibrating the system far above the freezing temperature, or a highly heterogeneous structure with domains enriched in some components by rapid freezing of metastable clusters formed near the freezing temperature.

Finally, we would like to address some of the limitations of the present work. While we believe that most of our qualitative results would apply to three dimensions as well, the question of whether the critical-like behavior observed in 2d will carry over to 3d (where the freezing transition is definitely 1st order in one component systems) do to the peculiarities of APD/AID systems, remains open. Such a study would require working with much larger systems in order to minimize finite size effects that tend to broaden the transition. Another potential problem of using the unbounded E distribution is that bonds with high enough $\epsilon_{ij}$ may not be equilibrated in our simulations and this may result in an AID glass whose properties depend on the particular choice of interaction parameters from the exponential distribution [11], even at temperatures above the freezing transition. Although such effects were not systematically explored in the present work, we checked that different realizations of the set $\{\epsilon_{ij}\}$ yield the same $< \epsilon_{ij}(T) >$, for temperatures above freezing. We believe that the main effect of such bonds is to form stable "diatomic" molecules which are accounted for in our simulations (Fig. 10b).

## Supplementary Material

See Supplementary Material for (1) **Movie of E Clusters**, at $\delta = 0.02$ and $t_{cutoff} = 100 \, \tau_{LJ}$. The 5 largest clusters are shown in brown. The rest of the particles are in blue.

Acknowledgments

This work was supported by grants from the Israel Science Foundation and from the Israeli Centers for Research Excellence program of the Planning and Budgeting Committee. Helpful comments by Dino Osmanovic are gratefully acknowledged.